\documentclass[10pt,twoside,natbib]{article}
\usepackage{epsfig,pkas,natbib}
\makehead{S. C. KIM ET AL.}{{\it Nature} and {\it Science} Papers}

\def\simlt{\lower.5ex\hbox{$\; \buildrel < \over \sim \;$}}
\def\simgt{\lower.5ex\hbox{$\; \buildrel > \over \sim \;$}}

\begin{document}
\sloppy
\pagenumbering{arabic}

\twocolumn[
\pkastitle{27}{1}{13}{2012}
\begin{center}
{\large \bf 
Trends of Papers Published from 2006 to 2010 in Journals {\it Nature} and {\it Science}} \vskip 0.5cm
{\bf }
{\sc Sang Chul Kim$^1$, Seung-Lee Kim$^1$, Jaemann Kyeong$^1$, Jae Woo Lee$^1$,
     Wanggi Lim$^1$, and Sung Hyun Jeon$^{1,2}$} \\
$^1$Korea Astronomy and Space Science Institute, Daejeon 305-348, Republic of Korea \\
$^2$Department of Astronomy and Space Science, Chungnam National
  University, Daejeon 305-764, Republic of Korea \\
\normalsize{\it (Received ???. ??, 2012; Accepted ???. ??, 2012)} \\
{\it E-mail: sckim@kasi.re.kr, slkim@kasi.re.kr, jman@kasi.re.kr, 
  jwlee@kasi.re.kr, wlim@kasi.re.kr, pupi0040@gmail.com}
\end{center}
\newabstract{
We present an analysis of the papers published in the journals 
  {\it Nature} and {\it Science} in the years from 2006 to 2010.
During this period, a total of 7788 papers were published 
  in the two journals.
This includes 544 astronomy papers that comprise 
  7.0\% of the papers in `all' research fields
  and 18.9\% of those in the fields of `physical sciences'.
The sub-fields of research of the astronomy papers
  are distributed,
  in descending order of number of papers, 
  in Solar System, stellar astronomy, galaxies and the universe,   
  the Milky Way Galaxy, and exoplanets.
The observational facilities used for the studies are mainly 
  ground-based telescopes (31.1\%), spacecrafts (27.0\%), and
  space telescopes (22.8\%), 
  while 16.0\% of papers did not use any noticeable facilities and
  1.7\% used other facilities.
Korean scientists have published 
  86 papers (33 in {\it Nature} and 53 in {\it Science}),
  which is 1.10\% of all the papers (N=7788) in the two journals.
The share of papers by Korean astronomers among the scientific papers by Koreans
  is 8.14\%, slightly higher than the contribution of astronomy papers 
  (7.0\%) in both journals.
\vskip 0.5cm {\em key words:}
history and philosophy of astronomy
 --- sociology of astronomy
 --- astronomical data bases: miscellaneous
} \vskip 0.15cm ] \flushbottom 

\newsection{Introduction}
\label{sect:intr}

While all scientific and astronomical research papers report new findings
  on nature and the Universe,
  some of them contribute greatly to the development of science
  and human knowledge.
These achievements that have a high impact on science and mankind
  are often acknowledged by press releases to the public
  and/or prestigious prizes such as the Nobel Prize.
Some representative ways to assess papers and
  determine which are high impact papers
  could be investigating highly cited papers or
  considering papers published in highly cited journals
  \citep{davoust87, leverington96, schulman97, abt98, abt00, pearce04,
  trimble08, stanek08, crabtree08, trimble09, frogel10, kim11}.

There have been many studies to measure the productivity and/or
  effectiveness of (1) facilities (e.g. telescopes) 
  \citep{trimble95, trimble96, benn01a, benn01b, ringwald03, 
  meylan04, trimble05, grothkopf05, trimble07, trimble08, trimble09, apai10}, 
  (2) organizations \citep{crabtree03},
  (3) countries \citep{sanchez04, abt10, kamphuis10}, 
  (4) scholars \citep{dietrich08,stanek09,kamphuis10, pimbblet11}, and so on.
\citet{ahn08} suggested that the number of papers produced by
  ground-based large ($D \sim 3.6 - 10$ m) optical telescopes are roughly
  proportional to the diameters of the primary mirrors
  (see also \citet{leverington97}).
Recently, \citet{kim11} presented results of an investigation on 
  the paper productivities of ground-based large ($D> 8$ m)
  optical telescopes from an analysis of papers published from 2000 to 2009.

Considering that the astronomical papers with the highest number of citations
  and those published in the journals {\it Nature} and {\it Science} 
  are the outputs with the greatest impact on science and on mankind
  \citep{benn01b},
  we have investigated the papers 
  published in the journals {\it Nature} and {\it Science} 
  from 2006 and 2010 in this study.
Specifically, we have tried to answer the following questions
  which people often ask: (1) how many papers are published in the journals 
    {\it Nature} and {\it Science}, 
  (2) what is the percentage of astronomy papers among these papers,
  (3) what are the distributions and portions of sub-research fields of astronomy,
  (4) what facilities were used for those astronomy papers and 
    what were the percentages of their uses, and
  (5) how many Korean scientists and Korean astronomers contributed to those
    papers. 
Because in some countries including Korea there is insufficient capability
  in the society
  to assess the scientific competence of personnel or the qualities of 
  research output, it is common to consider publications in
  highly cited journals/magazines like {\it Nature} and {\it Science}
  as the proxy of scientific expertise.
It will be, therefore, meaningful to investigate the statistics and
  distribution of papers in the two representative journals.
This paper is organized as follows:
  Section 2 describes the data utilized in this work.
Section 3 presents the analysis results of the number of papers,
  research fields in astronomy, astronomical facilities used, and
  papers by Koreans.
Finally, Section 4 provides summary and discussion of the results.

\newsection{Data}
\label{sect:data}

The academic papers investigated in this study are those contained in
  the two weekly journals 
  {\it Nature}\footnote{http://www.nature.com} and 
  {\it Science}\footnote{http://www.sciencemag.org} for five years from 2006 to 2010.
Among the contents of the two journals,
  we only counted `articles' and `letters' in {\it Nature} and
  `research articles' and `reports' in {\it Science}
  in order to take into account original studies 
  (cf. Isaac Newton Group webpage\footnote{http://www.ing.iac.es/PR/natsci.html}).

In this paper, we have used the term `astronomy' to include both
  astronomy and astrophysics.

\newsection{Results}
\label{sect:results}

\newsubsection{Number of Papers}
\label{sect:number}

Table \ref{tab1} shows the basic statistics of the papers in the journals
  {\it Nature} and {\it Science} from 2006 to 2010.
During this period, 4004 papers were published in {\it Nature},
  while 3784 papers were published in {\it Science},
  with yearly mean numbers of 800 and 757, and
  weekly (i.e., per issue) mean numbers of 15.6 and 14.8, respectively.

For the journal {\it Nature}, we have used the webpage 
  of the Japanese table of contents, which shows a detailed field classification
  for each article.
Using these classifications, we distributed each research field into
  two areas of life sciences and physical sciences,
  of which items are shown in the footnotes of Table \ref{tab1}.
The second column of Table \ref{tab1} shows the yearly number of papers 
  in each of these two main categories for the journal {\it Nature}.

The subject index of the journal {\it Science}'s webpage gives three main classifications;
  life sciences, physical sciences, and other subjects.
Astronomy is included in the physical sciences;
  `other subjects' include education, economics, sociology, policy/research ethics,
  etc.
The numbers of papers in each of these categories for each year are 
  shown in the fourth column of Table \ref{tab1}.

For the period of 2006 to 2010,
  there were 319 and 225 astronomy papers published
  in the journals {\it Nature} and {\it Science},
  with yearly mean values of 64 and 45, respectively.
The astronomy papers in {\it Nature} accounted for 8.0\% and 24.1\%
  of `all' and `physical sciences' papers, 
  while those of {\it Science} accounted for 5.9\% and 14.4\%, respectively.
These 544 ($=319+225$) papers in astronomy for the journals 
  {\it Nature} and {\it Science} comprise 
  a total of 7.0\% of the papers for 'all' research fields (N$= 7788 =4004+3784$) and
  18.9\% of the papers for the fields of 'physical sciences' (N$=2885 =1321+1564$).

If we simply compare the fraction of astronomy papers among all science papers 
  with the fraction of funds given to astronomy field among all research fields,
the portion of astronomy papers among all science papers is greater than
  that of the fund given to astronomy among all the research related budgets.
For example, in the case of the United States (we take the U.S. case
  as an example because it is not easy to get funding information
  for other countries),
  funds approved and disbursed by the National Science Foundation 
  for the field of astronomy in 2011 
  were only 4.3\% (236.6 million USD) 
  of the research and related activities (R\&RA) fund of 
  5.56 billion USD\footnote{http://www.nsf.gov}.
All the more interesting is to note that astronomy papers make up
  almost one fifth of the physical science papers,
  which include all the natural sciences and engineering fields.
This implies that astronomy has a
  greater impact on science and mankind 
  and generates great interest from the general public.

\begin{table*} 
\small{  
\begin{center}
\caption{Number of Papers for Research Fields Published in 
  the {\it Nature} and {\it Science} Journals}\label{tab1}
\begin{tabular}{cllll}
\hline\hline
{Year} & \multicolumn{2}{l}{\it Nature} & \multicolumn{2}{l}{\it Science} \\
       & N(all)$^{\rm a}$ & N(astronomy)$^{\rm b}$ & N(all)$^{\rm c}$ & N(astronomy)$^{\rm b}$ \\
 (1)   & (2)              & (3)                    & (4)              & (5) \\
\hline
2006 & 822 $(=535+287)$    & 82 (10.0\%, 28.6\%) & 758 $(=434+313+11)$   & 41 (5.4\%, 13.1\%)\\
2007 & 762 $(=512+250)$    & 50  (6.6\%, 20.0\%) & 743 $(=431+303+ 9)$   & 34 (4.6\%, 11.2\%)\\
2008 & 823 $(=555+268)$    & 65  (7.9\%, 24.3\%) & 748 $(=432+308+ 8)$   & 46 (6.1\%, 14.9\%)\\
2009 & 783 $(=515+268)$    & 62  (7.9\%, 23.1\%) & 774 $(=443+323+ 8)$   & 48 (6.2\%, 14.9\%)\\
2010 & 814 $(=566+248)$    & 60  (7.4\%, 24.2\%) & 761 $(=437+317+ 7)$   & 56 (7.4\%, 17.7\%)\\
\\
Sum  &4004 $(=2683+1321)$  &319  (8.0\%, 24.1\%) &3784 $(=2177+1564+43)$ &225 (5.9\%, 14.4\%)\\
     & 100 $(=67.0+33.0)$\%&                     & 100 $(=57.5+41.3+1.1)$\% & \\
\hline
\end{tabular}
\medskip\\
\end{center}
$^{\rm a}$ Number of papers for all research fields
  $(={\rm life ~science}+{\rm physical ~science})$,
  where astronomy is included in the latter.
  {\bf Life sciences} means archaeology, structural biology, agronomy, nano-technology, brain,
  immunology, microbiology, virus, ontogeny, pathology,
  taxonomy, physiology,
    biophysics, ecology, biochemistry,
  cytology, sense, plant, neurology, psychology,
  cancer, pharmacology, linguistics, epidemiology, genetics,
  cognizance, medicine, tumor, evolution, and cooperative action 
  and {\bf physical sciences} means
  engineering, nano-technology [different from that
    in the life science], mathematics, universe meteorology,
  physics, physical chemistry, optics, quantum, materials,
  earth, statistics, microscopy, ocean, visualization,
  chemistry, chemical engineering, and environment. \\
$^{\rm b}$ Number of papers in astronomy. Values in parentheses are 
  percentage among `all' papers and
  percentage among `physical science' papers, respectively. \\
$^{\rm c}$ Number of papers for all research fields $(={\rm life ~science}+{\rm physical ~science}+
  {\rm Etc.})$, where 
  `Etc' includes, e.g., education, economics, sociology, and policy/research ethics 
  (astronomy is included in the `physical science') \\
} 
\end{table*}

\newsubsection{Sub-fields in Astronomy}
\label{sect:fields}

Table \ref{tab2} shows the distribution of papers in
  the sub-fields of astronomy for the journals {\it Nature} and {\it Science}.
The classification criterion has been set up by the authors
  and modified by comparing the results from the two journals.
Figure 1 
  shows the general distribution of the sub-fields.
The data in this Figure shows that the order, sorted by number of papers, is 
  Solar System, stellar astronomy, galaxies and the universe, 
  the Milky Way Galaxy, and exoplanets.

`Solar System' is the most studied sub-field in the two journals,
  with a percentage of 37.9\%,
  followed by `Stars' (11.4\%),
  `External Galaxies' (10.5\%),
  `Supernovae and Novae' (7.2\%),
  `Exoplanets' (7.0\%), and so on (Table \ref{tab2}).
`Solar System' and `Stars' comprise half of astronomy papers (see the sixth column
  of Table \ref{tab2}),
  while the five fields (Solar System, Stars, External Galaxies, Supernovae and Novae,
  and Exoplanets) make up three quarters.

Reasons why the field of `Solar System' takes the largest portion of astronomy papers
  could be the following.
The first reason could be the launches of several spacecrafts/satellites,
  which bring us much closer and more detailed views/information
  on Solar System objects.
This is shown in the following subsection, in Table \ref{tab3} and in Figure 2. 
As can be seen in Figure 2 (a), the papers that used spacecrafts comprised
  27.0\% (147/544) of all the astronomy papers; 
  and spacecrafts were used in 32.2\% (147/457) 
  of astronomy papers (excluding papers of `no facility used').
Another reason might be the great interest of both scientists and the public
  in the neighborhood of our home planet, which extends from the Earth and Moon
  to Mercury, Venus, and Mars, and on to the far side of the Solar System,
  as well as to asteroids and comets.

\begin{table*} 
\small{  
\begin{center}
\caption{Sub-field Distribution of Astronomical Papers in the {\it Nature} 
  and {\it Science} Journals from 2006 to 2010$^{\rm a}$}\label{tab2}
\begin{tabular}{cccccc}
\hline\hline
{Field} & {N({\it Nature})} & {N({\it Science})} & {N(Sum)} & {Percentage [\%]} &
  {Accumulated} \\
&&&&& {Percentage [\%]} \\
(1) & (2) & (3) & (4) & (5) & (6) \\
\hline
Solar System                                       &107 &99 &206 &37.9 & 37.9 \\
Stars                                              & 32 &30 & 62 &11.4 & 49.3 \\
External Galaxies                                  & 45 &12 & 57 &10.5 & 59.7 \\
Supernovae and Novae                               & 23 &16 & 39 & 7.2 & 66.9 \\
Exoplanets                                         & 28 &10 & 38 & 7.0 & 73.9 \\
Formation of Stars and the Solar System            & 13 &14 & 27 & 5.0 & 78.9 \\
Interstellar Matter (including Supernova Remnants) & 11 &12 & 23 & 4.2 & 83.1 \\
Gamma Ray Bursts                                   & 17 & 5 & 22 & 4.0 & 87.1 \\
Cosmology                                          & 10 & 4 & 14 & 2.6 & 89.7 \\
ilky Way Galaxy                                   &  8 & 4 & 12 & 2.2 & 91.9 \\
Star Clusters                                      &  7 & 4 & 11 & 2.0 & 93.9 \\
Sun                                                &  6 & 5 & 11 & 2.0 & 96.0 \\
Active Galactic Nuclei                             &  7 & 3 & 10 & 1.8 & 97.8 \\
Galaxy Clusters and Large Scale Structure          &  3 & 5 &  8 & 1.5 & 99.3 \\
Cosmic Ray                                         &  1 & 2 &  3 & 0.6 & 99.9 \\
Instrumentation                                    &  1 & 0 &  1 & 0.2 & 100  \\
&&&& \\
Sum                                               & 319 & 225 & 544 & 100 & -- \\
\hline
\end{tabular}
\medskip\\
\end{center}
$^{\rm a}$ On the order of percentage \\
} 
\end{table*}

\begin{figure*}
\epsfxsize=13.0cm 
\centerline{\epsffile{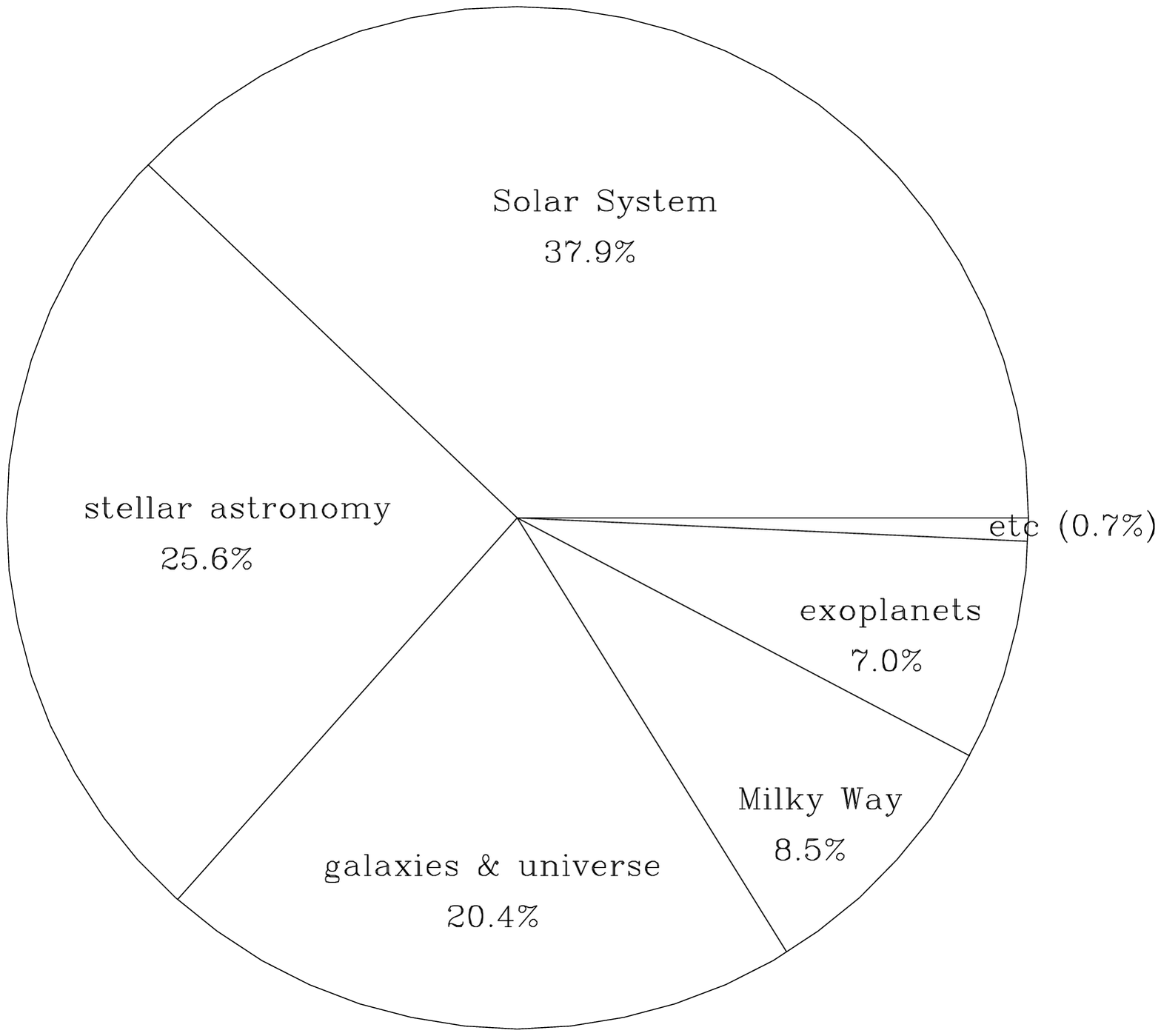}}
{\small {\bf ~~~Fig. 1.}---~The distribution of subjects 
 for papers published in the journals {\it Nature} and {\it Science}.
`Stellar astronomy' includes stars, supernovae, novae, star formation, and
  the Sun; `galaxies \& universe' includes external galaxies, active galactic
  nuclei, gamma-ray bursts, galaxy clusters, large scale structure, and
  cosmology; `Milky Way' includes our Galaxy, star clusters, and
  interstellar matter; `etc.' includes cosmic rays and instrumentation.
This shows only the general trend and the exact order could change,
  depending on the classification criteria of each item.
}
\label{fig1}
\end{figure*}

\newsubsection{Facilities for Astronomy Papers}
\label{sect:facilities}

Table \ref{tab3} and Figure 2 
  show the statistics of the observational facilities used 
  in the astronomy papers under consideration.
When multiple facilities are used in a certain paper,
  we either (1) focused on the main facility which is presumed to have provided
  the most critical data for the research
  (e.g., taking the Very Large Telescope instead of the Keck Telescope 
  in \citet{galyam09}'s paper),
  or (2) took the larger (more expensive) facility over the smaller(cheaper) ones 
  when they were used
  with similar importance (e.g., taking the 8.2 m Subaru Telescope
  instead of the 2.2 m/3.5 m telescopes at the Calar Alto Astronomical Observatory,
  in \citet{krause08}).
Although these selections can leave some ambiguities in certain cases,
  we assume that they do not significantly affect the distribution
  shown in Figure 2. 

While 16.0\% of papers did not use any noticeable facilities 
  for their studies (e.g., theory, simulation),
  large (D $> 8$ m) optical telescopes were dominantly used 
  for the studies of 10.5\% papers (combining {\it Nature} and {\it Science}).
They are currently the largest facilities in optical wavebands.
Table \ref{tab3} shows that
  the next most heavily-used facilities were spacecrafts to Saturn and Mars (14.9\%) and
  space telescopes in the gamma-ray, optical, and infrared wavebands (15.8\%).
Figure 2 (a) shows that the percentages of papers that used
  ground-based telescopes, spacecrafts, and space telescopes were
  31.1\%, 27.0\%, and 22.8\%, respectively, for all astronomy papers.
Optical and radio \& sub-mm telescopes make up
  69.2\% and 23.7\%, respectively, of all the ground-based telescopes
  (Figure 2 (b)). 
Figure 2 (c) shows that space telescopes of
  gamma-ray, optical, infrared, X-ray, and ultraviolet wavebands
  take 24.2\%, 23.4\%, 21.8\%, 16.1\%, and 3.2\%, 
  respectively. 

Among the 105 papers that used ground-based optical telescopes, 
  as shown in Figure 2 (d),
  the largest telescopes of D $>$ 8 m comprise 54.3\% (N=57), while
  those of 3.0 $\leq$ D $\leq$ 4.2 m take 21.0\% (N=22),
  those of D $<$ 3.0 m take 19.0\% (N=20) and
  those of 5.0 $\leq$ D $\leq$ 6.5 m take 5.7\% (N=6).
The possible reason why ground-based optical telescopes of diameters 
  between 5.0 m and 6.5 m take a lesser percentage of papers (1.1\% among all 
  the facilities in Table \ref{tab3}) than
  those of diameters between 3.0 m and 4.2 m (4.0\%) or
  even those of diameters smaller than 3.0 m (3.7\%) (Figure 2 (d)) could be 
  the lower ($\simlt 1/3$) number of telescopes for ground-based optical telescopes
  of diameters between 5.0 m and 6.5 m.
As can be seen in Table \ref{tab3},
  among the ground-based optical telescopes, it is remarkable that
  the Sloan Digital Sky Survey (SDSS) project,
  operated with only one 2.5 m telescope,
  stands out with 2.2\% of papers.
This might result from a large survey program covering a quarter
  of the sky and the creation of 3-dimensional maps containing more than 930,000
  galaxies and more than 120,000 quasars via both photometry and spectroscopy
  \citep{abazajian09}.
Although the large fraction of ground-based optical telescopes of D $>$ 8 m 
  (Figure 2 (d)) could be somewhat biased, if at all,
  by the second criterion explained in the first paragraph of this subsection,
  it is still valuable to note that state-of-the-art facilities
  and big science
  (e.g. space telescopes, ground-based optical telescopes of D $>$ 8 m) and
  dedicated facilities (e.g. SDSS, Cassini, Voyager, CoRoT)
  might be one of the critical factors to create high impact papers.

\begin{table*} 
\small{  
\begin{center}
\caption{Facilities Used in the Papers of {\it Nature} and {\it Science} Journals
  from 2006 to 2010}\label{tab3}
\begin{tabular}{lcccc}
\hline\hline
{Facility} & {\it Nature} & {\it Science} & Sum (\%) & {Accumulated} \\
&&&& {Percentage [\%]$^{\rm a}$} \\
(1) & (2) & (3) & (4) & (5) \\
\hline
no facility used                              &58      &29      &87 (16.0) &  -- \\
ground-based telescope, optical (D $>$ 8 m)   &39      &18      &57 (10.5) & 10.5\\
spacecraft - Cassini                          &28      &18      &46 (8.5)  & 18.9\\
spacecrafts to Mars                           &12      &23      &35 (6.4)  & 25.4\\
space telescope, gamma-ray                    &15      &15      &30 (5.5)  & 30.9\\
space telescope, optical                      &24      &5       &29 (5.3)  & 36.2\\
space telescope, infrared                     &18      &9       &27 (5.0)  & 41.2\\
radio telescope                               &13      &11      &24 (4.4)  & 45.6\\
ground-based telescope, optical (3.0 $\leq$ D $\leq$ 4.2 m) &19      &3       &22 (4.0)  & 49.6\\
spacecrafts to Moon                           &2       &20      &22 (4.0)  & 54.0\\
ground-based telescope, optical (D $<$ 3.0 m) &16      &4       &20 (3.7)  & 57.3\\
space telescope, x-ray                        &13      &7       &20 (3.7)  & 61.0\\
SDSS                                          &8       &4       &12 (2.2)  & 63.2\\
VLA, VLBA, VLBI                               &5       &7       &12 (2.2)  & 65.4\\
spacecraft - Others                           &4       &8       &12 (2.2)  & 67.6\\
spacecraft - Voyager                          &8       &2       &10 (1.8)  & 69.5\\
ground-based telescope, gamma-ray             &3       &5       &8 (1.5)   & 70.9\\
supercomputer                                 &5       &3       &8 (1.5)   & 72.4\\
spacecrafts to Mercury                        &0       &7       &7 (1.3)   & 73.7\\
spacecrafts to Earth                          &1       &6       &7 (1.3)   & 75.0\\
ground-based telescope, optical (5.0 $\leq$ D $\leq$ 6.5 m) &5       &1       &6 (1.1)   & 76.1\\
CoRoT satellite                               &4       &2       &6 (1.1)   & 77.2\\
Lunar sample/meteorite                        &5       &1       &6 (1.1)   & 78.3\\
submm telescope                               &2       &2       &4 (0.7)   & 79.0\\
space telescope, ultraviolet                  &3       &1       &4 (0.7)   & 79.8\\
space telescope, Sun observing                &1       &2       &3 (0.6)   & 80.3\\
balloon                                       &2       &0       &2 (0.4)   & 80.7\\
ground-based telescope, Sun observing         &0       &2       &2 (0.4)   & 81.1\\
ground-based telescope, cosmic ray            &0       &2       &2 (0.4)   & 81.4\\
space telescope - Kepler                      &0       &2       &2 (0.4)   & 81.8\\
spacecrafts to Venus                          &1       &1       &2 (0.4)   & 82.2\\
space telescope - WMAP                        &0       &1       &1 (0.2)   & 82.3\\
etc$^{\rm b}$                                 &5       &4       &9 (1.7)   & 84.0\\
\\
Total                         &319     &225     &544 (100) & --\\
\hline
\end{tabular}
\medskip\\
\end{center}
$^{\rm a}$ excludes the first row of `no facility used' (16.0\%). Accumulation
  starts from the second item (optical telescope, D$>$8 m) and the final sum
  becomes 84.0\%. \\
$^{\rm b}$ includes virtual observatory, Center for High Angular Resolution Astronomy
  (CHARA interferometer), Navy Prototype Optical Interferometer (NPOI), 
  dark matter search detector array, Laser Interferometer Gravitational-wave Observatory (LIGO), 
  composition analyzer, velocimeter, magnetometer \\
} 
\end{table*}

\begin{figure*}
\epsfxsize=17.0cm
\centerline{\epsffile{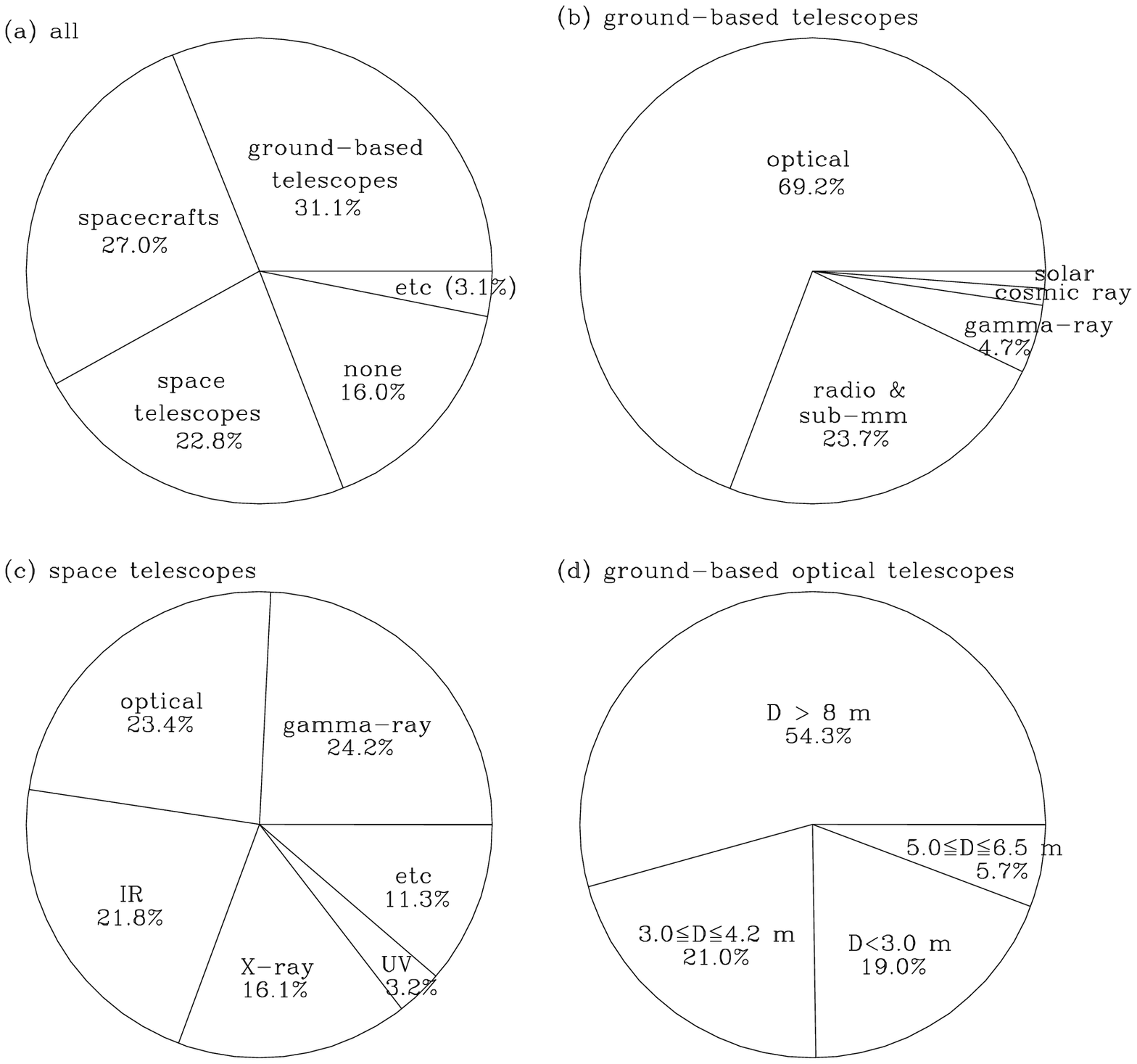}}
{\small {\bf ~~~Fig. 2.}---~Pie charts for the facilities used 
  in astronomy papers published in the journals
  {\it Nature} or {\it Science} from 2006 to 2010.
(a) Facilities for all 544 astronomy papers.
Spacecrafts include Lunar sample/meteorite;
  space telescopes include balloon and CoRoT satellites; 
  `etc.' includes supercomputer and the last item in Table \ref{tab3}.
(b) Sub-distribution of ground-based telescopes for all wavelength ranges.
`Optical' includes the SDSS telescope, and
  `radio \& sub-mm' includes VLA, VBLA, and VLBI telescopes.
(c) Sub-distribution of space telescopes, where
  `etc.' contains solar telescopes, Kepler, WMAP, balloons, and the CoRoT satellite.
(d) Sub-distribution of ground-based optical telescopes.
}
\label{fig2}
\end{figure*}

\newsubsection{Papers by Koreans}
\label{sect:koreans}

For the papers published in {\it Nature} and {\it Science} from 2006 and 2010,
  we have probed the papers with Korean scientists in the author lists
  and show the results in Table \ref{tab4}.
We extracted papers with authors of Korean names and Korean affiliations.

Table \ref{tab4} shows that 
  86 papers (33 for {\it Nature} and 53 for {\it Science}) have 
  Koreans as the authors, and among them
  seven papers (two for {\it Nature} and five for {\it Science})
  are astronomy papers written by Korean astronomers.
In total, Korean authors contributed 
  1.10\% of the total 7788 papers 
  published in the two journals.
Astronomy papers by Korean authors (N$=7$) 
  make up 8.14\% of the 86 papers by Korean scientists.
While astronomy papers comprise 7.0\% of the papers of 'all' research fields
  in the two journals {\it Nature} and {\it Science}, as can be seen 
  in Section 3.1,
  this contribution of Korean astronomers to the Korean sciences 
  (at least in the statistics of the two journals for the given period)
  shows a slightly higher percentage than the world normal.
If we take only the first author and corresponding author papers,
  the rate even increases to 11.1\% (5/45).

Table \ref{tab5} shows the number distribution of papers written by Korean scientists
  in the fields of physics, chemistry (including biochemistry), 
  biology \& life science, earth sciences \& astronomy, and 
  engineering (including materials science).
Although biology \& life science is the field of most abundance and 
  earth sciences \& astronomy is the field with least number of papers,
  the fraction of papers in biology \& life science in Korea (29.1\%) is 
  much lower than the fraction of this field in the world (Table 1; 
  4860/7788 = 62.4\%).
It is also worthwhile to note that astronomy papers make up
  two thirds (7/11) of the papers in the field of earth sciences \& astronomy.

Table \ref{tab6} provides a detailed bibliography of the seven
  {\it Nature} and {\it Science} papers
  written by Korean astronomers from 2006 and 2010,
  of which five papers are with first/corresponding authors
  and the remaining two are with co-authors.
Among the seven papers, two \citep{yoon06, ryu08} used no observational facilities;
  two papers that used GALEX\footnote{Galaxy Evolution Explorer satellite} as the main 
    facility \citep{schawinski06, schawinski08} actually used multiple facilities;
  one paper \citep{lee10} used the SDSS data; and
  the remaining two papers \citep{lee09, gaudi08} used small optical telescopes
  of 1 m class.
This status of facilities used by Korean astronomers reflects well
  the current situation of facilities for the Korean astronomical community,
  and shows (1) participation in one space project (GALEX),
  (2) use of small ground-based optical telescopes (CTIO 1.0 m and Mt. Lemmon
  1.0 m telescopes), (3) use of public archive data (SDSS), and 
  (4) studies without any noticeable facilities.
Since the tools that we use to look at the Universe are essential in astronomical studies,
  as can be seen in the Table \ref{tab3},
  construction of or participation in more facilities/projects will bring
  a greater number of {\it Nature} and {\it Science} papers in the future.

While the sub-fields of the seven {\it Nature} and {\it Science} papers
  published by Korean astronomers are diverse,
  it is interesting that three (43\%) out of seven papers
  are on `star clusters'.
It is true, in general, that scientists with more and better facilities 
  produce more and better papers.
Nevertheless, considering the fact that these three papers did not use
  any of the current largest (or most expensive) facilities,
  this statistic shows that excellence in scholarship
  is another indispensable element in its own way.

\begin{table*} 
\small{  
\begin{center}
\caption{Papers Written by Korean Scientists$^{\rm a}$ from 2006 to 2010}\label{tab4}
\begin{tabular}{ccccccc}
\hline\hline
{Journal} & \multicolumn{3}{l}{All Fields$^{\rm b}$} & \multicolumn{3}{l}{Astronomy} \\
          & N & (=First Author$^{\rm c}$ + & Others)& N &(=First Author$^{\rm c}$ + & Others)\\
\hline
{\it Nature}  & 33 &(=17 +   & 16)  & 2 &(=2 + & 0) \\
              & 0.82\%$^{\rm d}$ &(=0.42\%$^{\rm d}$+ & 0.40\%$^{\rm d}$)& 6.06\%$^{\rm e}$ & & \\
& & & & & &\\
{\it Science} & 53 & (=28 +   & 25) & 5 &(=3 + & 2) \\
              & 1.40\%$^{\rm f}$ &(=0.74\%$^{\rm f}$+ & 0.66\%$^{\rm f}$)& 9.43\%$^{\rm g}$ & & \\
\\
Sum           & 86 & (=45+    & 41) & 7 &(=5 + & 2) \\
              & 1.10\%$^{\rm h}$ &(=0.58\%$^{\rm h}$+ & 0.53\%$^{\rm h}$)& 8.14\%$^{\rm i}$ & & \\
\hline
\end{tabular}
\medskip\\
\end{center}
\hskip 1cm $^{\rm a}$ Korean names and Korean affiliations\\

\hskip 1cm $^{\rm b}$ Including astronomy \\

\hskip 1cm $^{\rm c}$ Including Corresponding Author \\

\hskip 1cm $^{\rm d}$ Percentage among the 4004 {\it Nature} papers from 2006 to 2010\\

\hskip 1cm $^{\rm e}$ Percentage among the 33 {\it Nature} papers by Korean scientists from 2006 to 2010\\

\hskip 1cm $^{\rm f}$ Percentage among the 3784 {\it Science} papers from 2006 to 2010\\

\hskip 1cm $^{\rm g}$ Percentage among the 53 {\it Science} papers by Korean scientists from 2006 to 2010\\

\hskip 1cm $^{\rm h}$ Percentage among the 7788 {\it Nature} and {\it Science} papers from 2006 to 2010\\

\hskip 1cm $^{\rm i}$ Percentage among the 86 papers by Korean scientists from 2006 to 2010\\
} 
\end{table*}

\begin{table*} 
\small{  
\begin{center}
\caption{Number of Papers Written by Koreans$^{\rm a}$ in Each Field}\label{tab5}
\begin{tabular}{cllllc}
\hline\hline
          & \multicolumn{2}{l}{\it Nature} & \multicolumn{2}{l}{\it Science} & Sum(\%) \\
{Field}          & First Author$^{\rm b}$ & Others & First Author$^{\rm b}$ & Others & \\
\hline
Physics                     & 6 & 6 &  3 & 7 & 22 (25.6\%)\\
Chemistry                   & 4 & 1 &  4 & 5 & 14 (16.3\%) \\
Biology \& Life Science     & 1 & 7 & 12 & 5 & 25 (29.1\%) \\
Earth Science \& Astronomy  & 3(2)$^{\rm c}$ & 0 &  4(3)$^{\rm c}$ & 4(2)$^{\rm c}$ & 11 (12.8\%) \\
Engineering                 & 3 & 2 &  5 & 4 & 14 (16.3\%) \\
\hline 
Sum                       & 17 & 16 & 28 & 25 & 86 (100\%) \\
\hline
\end{tabular}
\medskip\\
\end{center}
\hskip 1cm $^{\rm a}$ Korean names and Korean affiliations\\

\hskip 1cm $^{\rm b}$ Including Corresponding Author \\

\hskip 1cm $^{\rm c}$ The number in parentheses is that for astronomy \\
} 
\end{table*}

\begin{table*} 
\small{  
\begin{center}
\caption{Bibliography of Papers Written by Korean Astronomers 
  from 2006 to 2010$^{\rm a}$}\label{tab6}
\begin{tabular}{lll}
\hline\hline
1$^{\rm b}$ & Authors & Schawinski, Kevin; Khochfar, Sadegh; Kaviraj, Sugata; Yi, Sukyoung K.; 15 coauthors ; \\
  & & \hskip 5mm  Lee, Young-Wook; and 4 coauthors \\
  & Journal & 2006, Nature, 442, 888 \\
  & Title & Suppression of star formation in early-type galaxies by feedback from supermassive black holes \\
  & Facilities & GALEX, SDSS \\
  & Subject & External Galaxies \\
& & \\
2$^{\rm c}$ & Authors & Lee, Jae-Woo; Kang, Young-Woon; Lee, Jina; Lee, Young-Wook \\
  & Journal & 2009, Nature, 462, 480 \\
  & Title & Enrichment by supernovae in globular clusters with multiple populations \\
  & Facilities & CTIO 1.0 m \\
  & Subject & Star Clusters \\
& & \\
3$^{\rm c}$ & Authors & Yoon, Suk-Jin; Yi, Sukyoung Ken; Lee, Young-Wook \\
  & Journal & 2006, Science, 311, 1129 \\
  & Title & Explaining the Color Distributions of Globular Cluster Systems in Elliptical Galaxies\\
& Facilities & Models \\
& Subject & Star Clusters \\
& & \\
4$^{\rm c}$ & Authors & Ryu, Dongsu; Kang, Hyesung; Cho, Jungyeon; Das, Santabrata \\
  & Journal & 2008, Science, 320, 909 \\
  & Title &  Turbulence and Magnetic Fields in the Large-Scale Structure of the Universe \\
  & Facilities & Simulations\\
  & Subject & Large-scale structure \\
& & \\
5$^{\rm c}$ & Authors & Lee, Myung Gyoon; Park, Hong Soo; Hwang, Ho Seong \\
  & Journal & 2010, Science, 328, 334 \\
  & Title & Detection of a Large-Scale Structure of Intracluster Globular Clusters in the Virgo Cluster\\
  & Facilities & SDSS \\
  & Subject & Star Clusters \\
& & \\
6$^{\rm d}$ & Authors &Gaudi, B. S.; 19 coauthors ; 
 Han, C.; Kaspi, S.; Lee, C.-U.; 3 coauthors ;
 Park, B.-G.; \\
  & & \hskip 5mm and 47 coauthors\\
  & Journal & 2008, Science, 319, 927 \\
  & Title & Discovery of a Jupiter/Saturn Analog with Gravitational Microlensing \\
  & Facilities & LOAO and many telescopes \\
  & Subject & Exoplanets \\
& & \\
7$^{\rm d}$ & Authors & Schawinski, Kevin; 22 coauthors; 
Yi, Sukyoung K. \\
  & Journal & 2008, Science, 321, 223 \\
  & Title & Supernova Shock Breakout from a Red Supergiant \\
  & Facilities & GALEX, CFHT, VLT, Gemini, HST \\
  & Subject & Supernovae \\
\hline
\end{tabular}
\medskip\\
\end{center}
\hskip 2cm $^{\rm a}$ Names of non-Korean authors in the middle of author list
  are substituted with number of coauthors. \\

\hskip 2cm $^{\rm b}$ Paper with Korean Corresponding Author (Yi, Sukyoung K.)\\

\hskip 2cm $^{\rm c}$ First Author papers \\

\hskip 2cm $^{\rm d}$ Coauthor papers \\
} 
\end{table*}

\newsection{Summary and Discussion}
\label{sect:sum}

We have examined the distribution and statistics of
  `articles' and `letters' in the journal {\it Nature} and
  `research articles' and `reports' in the journal {\it Science}
  published from 2006 to 2010.
The 4004 {\it Nature} papers are composed of 2683 life science papers and
  the 1321 physical science papers, among which the latter group contains
  319 (8.0\% among 4004) astronomy papers.
The 3784 {\it Science} papers are made up of 2177 life science papers,
  1564 physical science papers, and 43 papers in other fields,
  where 225 (5.9\% among 3784) astronomy papers are included in the
  physical science papers.
In total, astronomy papers comprise 7.0\% of the papers for 'all' research fields
  and 18.9\% of the papers for the fields of 'physical sciences'
  in the two journals.

The sub-fields of study for these astronomy papers are as follows:
  `Solar System' (37.9\%) and `Stars' (11.4\%) comprise 
  half of the astronomy papers, while
  the five fields of Solar System, Stars, External Galaxies, Supernovae and Novae,
  and Exoplanets make up three quarters.

While 16\% of the astronomy papers did not use any noticeable facilities 
  for their research,
  spacecrafts, space telescopes, and ground-based telescopes were used
  for 27.0\%, 22.8\%, and 31.1\%, respectively.
Such spacecrafts, which explore in detail objects in the Solar System,
  might have been an important factor in increasing the large number of
  the research papers in this field.
Space telescopes are mainly those in the gamma-ray (24.2\%), optical (23.4\%), 
  infrared (21.8\%), and X-ray (16.1\%) wavebands; 
  ground-based telescopes are largely optical (69.2\%) and
  radio (23.7\%) telescopes.
The largest (D $>$ 8 m) ground-based optical telescopes produced the greatest 
  number of {\it Nature} and {\it Science} papers (57/105, 54.3\%) among all
  ground-based telescopes.
While this value could be affected somewhat, if not much, by one of the 
  facility-selection criteria in Section 3.3 (taking the larger facility
  as the primary facility for the paper
  when one large and one small facilities are used in equal amounts),
  it seems that the order of importance 
  among the ground-based optical telescopes would not change.

From the 4004 {\it Nature} and 3784 {\it Science} papers,
  we have extracted 86 papers by Korean authors with Korean
  affiliations. 
Among these 86 papers, seven astronomy papers (two in {\it Nature} 
  and five in {\it Science}) are included,
  making up 8.14\% of the 86 Korean papers.
While `astronomy' papers comprise 7.0\% of the papers for `all'
  the research fields of the journals {\it Nature} and {\it Science},
  Korean astronomers appear to contribute slightly more (8.14\%) to 
  all Korean papers in these two journals.

We anticipate that these results might be used, at least, for establishing criteria
  to assess leading research groups (especially for astronomy fields),
  and estimating future production of {\it Nature} and {\it Science} papers.

\section*{Acknowledgments} 
\vspace{4mm}
We would like to thank the anonymous referees for the very helpful comments and
  clarifications that helped to improve the manuscript. 
We would also like to thank Dr. Byeong-Gon Park for
  helping us with the NSF related information.


\end{document}